\documentclass[useAMS,usenatbib]{mn2e}
\usepackage{url}
\usepackage{amsmath}
\usepackage{algorithm}
\usepackage{algpseudocode}
\usepackage{color}
\usepackage{graphicx} % support the \includegraphics command and options
\renewcommand{\theenumi}{\arabic{enumi}.}

\title[Tera-scale Astronomical Data Analysis and Visualization]{Tera-scale Astronomical Data Analysis and Visualization}
\author[A. H. Hassan, C. J. Fluke, D. G. Barnes, and V. A. Kilborn]{A. H. Hassan$^{1}$\thanks{E-mail:
ahassan@swin.edu.au (AHH); cfluke@astro.swin.edu.au (CJF); david.g.barnes@monash.edu (DGB); vkilborn@astro.swin.edu.au (VAK)}, C. J. Fluke$^{1}$, D. G. Barnes$^{2}$, and V. A. Kilborn$^{1}$\\
$^{1}$Centre for Astrophysics and Supercomputing, Swinburne University of Technology, POBox 218, Hawthorn, Australia, 3122\\
$^{2}$Monash e-Research Centre, Monash University, Clayton, VIC 3800, Australia}
\begin{document}

\date{}

\pagerange{\pageref{firstpage}--\pageref{lastpage}} \pubyear{2012}

\maketitle

\label{firstpage}

\begin{abstract}
We present a high-performance, graphics processing unit (GPU)-based
framework for the efficient analysis and visualization of (nearly)
terabyte (TB)-sized 3-dimensional images.  Using a cluster
of 96 GPUs, we demonstrate for a 0.5 TB image: (1) volume rendering
using an arbitrary transfer function at 7--10 frames per second; (2)
computation of basic global image statistics such as the mean
intensity and standard deviation in 1.7 s; (3) evaluation of the image
histogram in 4 s; and (4) evaluation of the global image median intensity in just 45 s.  Our measured results correspond to
a raw computational throughput approaching one teravoxel per second, and
are 10--100 times faster than the best possible performance
with traditional single-node, multi-core CPU implementations.  A
scalability analysis shows the framework will scale well to images
sized 1 TB and beyond.  Other parallel data analysis algorithms can be
added to the framework with relative ease, and accordingly, we present
our framework as a possible solution to the image analysis and visualization requirements of
next-generation telescopes, including the forthcoming Square Kilometre
Array pathfinder radiotelescopes.

\end{abstract}


\section{Introduction}

\subsection{Radio Astronomy in the ``Big Data'' Era}
\label{subsection:bigdata}
Within the high-performance computing field, the term ``big data'' has been used to describe datasets too large to be handled with on-hand analysis, processing, and visualization tools. It is anticipated that the ability to perform these fundamental tasks will become a key basis for competition and science discoveries within the near future. Recent advances in astronomical observing and simulation facilities are expected to move astronomy toward a new data-intensive era where such ``big data'' is the norm rather than an exception. 

Upcoming radio telescopes such as the Australian Square Kilometre Array Pathfinder [ASKAP;\citet{johnston:2008}], MeerKAT array \citep{jonas2009meerkat}, and ultimately, the Square Kilometre Array (SKA)\footnote{\url{http://www.skatelescope.org/}} are clear examples of such facilities. While enabling astronomers to observe the radio universe at an unprecedented spatial and frequency resolution, handling the data from these facilities, expected to be from terabyte to petabyte order for individual observations, will pose significant challenges for current astronomical data analysis and visualization tools (e.g. \textsc{miriad}\footnote{\url{http://www.atnf.csiro.au/computing/software/miriad/}}, \textsc{kvis}\footnote{\url{http://www.atnf.csiro.au/computing/software/karma/}}, \textsc{ds9}\footnote{\url{http://hea-www.harvard.edu/RD/ds9/}}, and \textsc{casa}\footnote{\url{http://casa.nrao.edu/}}).

Data volumes that are orders of magnitude larger than astronomers and existing astronomy software are accustomed to dealing with will need revolutionary changes in data storage, transfer, processing, and analysis. We can summarize the limitations of current astronomical data analysis and visualization packages into the following:
\begin{itemize}
\item The majority of existing astronomical data analysis and processing solutions lack the ability to deal with datasets exceeding the local machine's memory limit, and so it will be a challenge to cope with such a massive increase in the data size.
\item Most data analysis systems are implemented as a set of separate tasks that can interact and exchange information via stored files only. This will be a significant factor which delays or even prohibits day-to-day data analysis tasks over tera-scale data sizes. Some exceptions to this norm exist. For example, the Platform for Astronomy Tool InterConnection (PLASTIC)\footnote{\url{http://www.ivoa.net/Documents/Notes/Plastic/PlasticDesktopInterop-20060601.html}} and the Simple Application Messaging Protocol (SAMP)\footnote{\url{http://www.ivoa.net/Documents/SAMP/}} have been developed within the virtual observatory framework to enable different client applications to communicate together (e.g. TOPCAT\footnote{\url{http://www.star.bris.ac.uk/~mbt/topcat/}} and Aladin\footnote{\url{http://aladin.u-strasbg.fr/}}). 
\item Some of the current data processing techniques depend on experimentally changing tuning-parameters  (e.g. thresholding based source finding, data smoothing, and noise removal), which might not be easy to achieve with such data sizes due to processing power limitations \citep{fluke:2010}.
\item It will no longer be an easy job to develop a simple script or program to deal with such data. Handling tera-scale datasets will involve development tasks that exceed the programming knowledge and experience available to the majority of astronomers. 
\end{itemize}

The work we present in this paper is one of the first in astronomy to address these limitations in a single solution (see \citet{kiddle2011cyberska} and \citet{kleijn2012data} for related approaches). We provide the ability for in-situ data analysis and visualization on big astronomical data. The ability to exchange information between different tasks in near realtime, and to work on the loaded dataset as an iterative pipeline process, will change the data analysis and processing model adopted by most astronomers. Moreover, the framework provides a relatively easy mechanism to add additional data analysis and processing functionality.   

Some facilities (e.g. ASKAP) are already planning an automated data processing pipeline that will work to perform data processing and information extraction tasks. However, due to operational requirements and limited processing capacity, such an automated pipeline will not be a silver-bullet solution for future data analysis and visualization demands. 
The main objective of this work is to prototype the role of quantitative and qualitative visualization tools in speeding-up different quality control, data analysis, and data processing tasks for tera-scale three-dimensional (3D) data.

\subsection{GPUs as an enabling technology}

A key problem in dealing with ``big data'' is the affordability of the computational resources required to analyze and visualize these datasets. The situation becomes more complicated when we add near real-time interactivity as an essential requirement. While most astronomical data is not time-sensitive, the requisite I/O capacity to store and retrieve such massive data volumes, and the continuous flow of the data that needs processing, will make interactivity and near-realtime processing critical capabilities. 

The massive floating point computational power of the modern general-purpose GPUs, and its relatively cheaper FLOPS/\$ and higher FLOPS/Watt compared to an equivalent CPU system, put it as one of the main players in the field of data intensive computing. We show through this work that GPUs combined with a distributed processing architecture are a solution for not only ``big data'' visualization but also for data analysis and processing tasks.

We will discuss the role of GPUs in our framework within the next sections in details. Giving a global review of GPUs and their role in astronomical data analysis and processing is outside the scope of this work. For a recent overview, see \citet{fluke:2011b}.

Previously, we have developed a distributed GPU-based framework for volume rendering large data [\citet{hassan:2011a} and \citet{hassan:2012a}]. In this work, we generalise this framework beyond qualitative volume rendering. The main contributions of this work are to:
\begin{itemize}
\item Introduce the Peer-to-Peer processing mode in addition to the client-server processing mode; 
\item Enable the user to use a generic (user controlled) transfer function in addition to the maximum intensity projection transfer function (see Section \ref{sct:TransferFunction} for details);
\item Introduce a new transfer function that emulates a local sigma clipping method to facilitate and speed-up the selection of sigma clipping thresholds; and
\item Introduce a sample of quantitative tools (e.g. 3D spectrum picking tool, distributed histogram calculation, and user controlled partitioning of the data  to enable local sigma clipping), which does not necessarily result in a visualisation product. These sample quantitative tools present a demonstration of how routine quantitative data analysis and  processing capabilities that are beyond all other existing analysis tools, can be implemented for terascale datasets and executed in just seconds on increasingly commonplace GPU-based clusters.
\end{itemize}

The paper is organized as follows. Section \ref{sct:DistributedGPUframework} discusses the framework's architecture and the design decisions taken through the framework design and implementation stages. Section \ref{sct:QualitativeVolumeRendering} describes the usage of our framework to implement distributed qualitative volume rendering for massive 3D datasets. Section \ref{sct:QuantitativeVisualization} describes the usage of our framework to integrate quantitative visualization and data analysis task with qualitative visualization.
Finally, Section \ref{sct:Results} shows the timing and performance measurements of different implemented processes and discusses the framework's scalability. 

\section[]{Distributed GPU framework}
\label{sct:DistributedGPUframework}
The main objective of the presented framework is to orchestrate different distributed nodes, each equipped with one or more GPUs, to work together to produce visualization output or to execute data processing tasks.
To provide an affordable and easy to use tera-scale data analysis and visualization tool, the following design decisions have been taken: 
\begin{itemize}
\item  To minimize the data movement, we chose to move the user questions to the data rather than moving the data to the user. We assume that the data is stored in a remote storage facility accessible by the GPU cluster. This enables our solution to work as a remote service by which the user can access data as a service and computing infrastructure as a service. This kind of service oriented architecture will hide the infrastructure complexity and provide a more cost-effective solution to a wider geographically distributed user community. 
\item Our solution assumes that it is going to work over a general-purpose  non-dedicated supercomputing facility (our system testing has utilized the The Commonwealth Scientific and Industrial Research Organisation GPU cluster\footnote{\url{http://www.csiro.au/Portals/Publications/Brochures--Fact-Sheets/GPU-cluster.aspx}} and the gSTAR Supercomputer at Swinburne University of Technology\footnote{\url{http://astronomy.swin.edu.au/supercomputing/green2/}}). 
\item We chose to use GPUs as the main processing element combined with the multi-core CPU to minimize the number of processing nodes required (with respect to a CPU only solution) and enable a near real-time processing and visualization. 
\item We adopt an in-core solution where the entire dataset is loaded into the GPU memory during the initialization stage. We chose the in-core solution to: avoid delays caused by the low-speed  disk I/O and data transformation from the main memory to the GPU memory within the data analysis and visualization; avoid the need to convert the currently widely adopted FITS file format\footnote{\url{http://fits.gsfc.nasa.gov/}} into another more distributed friendly data format; and support possible future memory-based integration with other tools as a data source or as a data destination.
\item Our solution mixes data analysis and visualization in the same framework so the user can see in real-time the effect of his operations or interact with the shown visualization results.   
\item Our implementation seeks to optimize data movement and memory usage more than optimizing the used processing power. With the existence of GPUs, the problem in hand is a memory limited problem rather than a processing limited problem. So, our algorithmic choices will take into consideration the computational complexity, but will give a higher weight to the memory needs and the data movement overhead (e.g. \citet{barsdell:2010}).
\end{itemize}
The main hardware architecture of our framework features three main components:
\begin{itemize}
\item \textbf{The remote client:} which acts as the main system interface and mainly works to interpret user graphical user interface (GUI) interactions into server commands and show the server output to the user within a suitable interactive GUI.
\item \textbf{The server:}  which acts as the interface between the main system backbone and the remote client. This provides the remote client with a single point of access, which facilitates securing the background cluster. 
\item \textbf{The GPU cluster:} which is the main processing backbone for the system.
\end{itemize}

To achieve its targets, the framework organizes the associated GPU processing nodes into two main modes based on the processing tasks:
\begin{itemize}
\item \textbf{Client-Server Processing mode} \\
In this mode, the different processing clients have no or limited communication between each other. The server is responsible for synchronizing the processing effort of each of them, does the data gathering, and merges the final result. The processing nodes use a direct-send communication pattern whereby they send the final processing results back to the server as soon as the processing is finished. This mode is quite useful when the result merging process does not require sophisticated computations, does not have big communication overhead, and does not depend on the order of the sub-results.

\item \textbf{Peer-to-peer Processing mode} \\
In this mode, the server is still responsible for the processing synchronization but the processing nodes work together to perform the result merging with no involvement of the server. The communication pattern between different processing nodes in this mode is dynamic and based on the data distribution and nature of the result merging task. This processing mode is useful when the merging processes require a specific sub-result order to ensure correctness, or when the communication messages' size and count are expected to cause  network congestion at the server node. 
\end{itemize}

It is the addition of the peer-to-peer processing mode that enables an efficient implementation of volume rendering with a generic transfer function, compared to our earlier solution, which used maximum intensity projection transfer function [\citet{hassan:2011a} and \citet{hassan:2012a}]. 

\section{Qualitative volume rendering for Quality control}
\label{sct:QualitativeVolumeRendering}

\subsection{From pretty pictures to interactive data processing}

The human brain is still the most intelligent data analysis, pattern recognition, and feature extraction tool. 
In a fraction of a second, our visual system can detect features that might require minutes from a supercomputer to identify. Current automated data analysis and processing tools can only detect known features and data characteristics. With the ``discovery of the unknowns'' as a major objective for future facilities, we think current automated data analysis and processing tools will not be a replacement for our human brain.  
In order to enable such discoveries, we need a method to summarize such massive datasets into a simple more easily interpreted form. That is one of the roles of scientific visualization. 

Volume rendering presents a better alternative to current, widely used, 2D dimensional techniques in providing global views of the data, particularly in the lack of clear feature segmentation \citep{hassan:2011a,hassan:2011b}. This global view is achieved by providing a pseudo-colour coded 2D projection(s) of discretely sampled points within the given 3D domain. The ability to generate a colour coded projection in an interactive manner (with higher than five frames per second), with the user controlling the projection viewport and the colour coding parameters, is a quite effective method to provide the user with global perspectives of the data. 

With respect to the available data representation and interaction tools, visualization services can be classified into two main categories:
\begin{enumerate}
\item \textit{Qualitative visualization}, which gives the user the global view of the data with the ability to interact with the displayed results and change different visualization parameters. This kind of visualization service is offered by the majority of available 3D visualization tools (some limitations related to the required computational power and maximum allowed data size exist). 
\item \textit{Quantitative visualization}, which adds to the qualitative visualization tools the ability to interrogate and further explore the data. 
\end{enumerate}
We think the lack of quantitative data analysis tools, and the relatively big computational demands for 3D visualization techniques, have contributed to the limited usage of 3D visualization tools and techniques in analysis and processing tasks. 

Interactive qualitative global data views can help the user to answer questions like: 
\begin{itemize}
\item Did something go wrong in the data gathering or processing?
\item Is the data good enough for my science purpose?
\item Do I need to change any tuning-parameters or request data re-processing?
\end{itemize}
The answer to these types of questions can provide essential feedback to the automated processing and reduction system. The second role is to help the user to provide inputs to future data analysis tasks, such as source finding [e.g. \citet{koribalski2012overview}]. Here, both qualitative and quantitative visualization can play an important role in helping astronomers to determine the noise characteristics, suitable noise removal technique and its parameters, and a useful local and global threshold values. The user can examine and see the outcome of changes in real-time, which can facilitate, speed-up, and enhance the quality of such data analysis tasks. We will discuss this further in Section \ref{sct:SigmaClippingTF}.

\subsection{Distributed Volume Rendering}
\label{sct:DistributedVolumeRenderingFramework}
To support distributed volume rendering across hybrid computing cluster comprising CPUs and GPUs, the system's software components are partitioned based on their memory boundaries into three main types: 

\begin{enumerate}
\item \textbf{The viewer} which is the GUI client responsible for controlling the volume rendering processes and displaying the final rendered output to the user; 
\item \textbf{The server} which acts as a middle layer between the viewer and the processing clients. The server here is a software component that can run as a separate thread on any of the processing nodes beside one or more processing clients. The server executes only simple CPU-based tasks; and 
\item \textbf{The processing client} component which is represented by two threads per client. The first thread controls the CPU operations and the communication with the server, while the second one controls the GPU operations. The usage of multi-threading enables asynchronous execution of the CPU and GPU tasks and helps to hide the communication overhead. 
\end{enumerate}
Usually, the number of processing clients is equal to the number of GPUs. Within this framework, each of the processing clients communicates with the server component or the other processing clients through MPI\footnote{\url{http://www.mcs.anl.gov/research/projects/mpi/}} messages as if they are not sharing any memory space. MPI automatically selects the fastest communication mechanism to send these messages (which for the processing client on the same node will be the shared memory).

\subsection{Transfer Functions}
\label{sct:TransferFunction}

The term ``transfer function'' is used to describe the mechanism of mapping data values into colours. The volume rendering algorithm is required to map a 3D data volume into a colour coded 2D projection. To do that, the transfer function is required to describe two main elements: how the data values along the line of sight of each output pixel are going to be merged and the mapping between the data values and the selected colour map. In this section, we discuss two main types of transfer functions: maximum intensity projection and generic transfer function. We will later discuss a newly introduced transfer function within this work, which we call the sigma-clipping transfer function (see Section \ref{sct:SigmaClippingTF}). 
 
\subsubsection{Maximum intensity projection}

The usage of the maximum intensity projection (MIP) has been addressed by us before in \citep{hassan:2011a,hassan:2012a}. The main advantage of using MIP as transfer function is its simplicity from the user perspective. It is a straight-forward  mapping method that requires minimum user involvement in determining the visualization parameters and can be easily utilized by users with limited scientific visualization experience. Furthermore, within the quality control context, the features the user is searching for (e.g large scale noise patterns or processing artifacts) are significant when they are much higher than their background. 

From the technical perspective, the overhead of merging rendered sub-frames is relatively lower because the maximum operator requires the processing nodes to exchange just the maximum value at each pixel, and it is an associative and commutative operator, which puts no restriction on using the client-server processing mode to implement it. On the other hand, only the value of the maximum voxel through each cast ray is shown for each pixel on the screen. By controlling the number of data levels in the colour-map and the minimum data value, the user can employ this method as a global thresholding technique to eliminate the noise or background information from the visualization output. 

Within this work, we re-implement this method using the peer-to-peer processing mode to put it in a unified implementation framework with the generic transfer function's volume rendering method [see Section \ref{sct:VolumeRenderingandFramecompositingprocesses} for details]. While the peer-to-peer processing mode increases the overall rendering time [with respect to the results published in \citet{hassan:2011a,hassan:2012a}], it is still within the interactive limit (which we define to be higher than five frames per second). 

\subsubsection {Generic transfer function volume rendering}

The MIP transfer function delivers a rendering that only represents one, outlier-based characteristic of the data (i.e. the maximal voxels). This representation omits the contribution of other data voxels to the rendering output in favour of simplicity. 
The generic transfer function is a more sophisticated data-to-colour mapping technique. The data values' range is partitioned into discrete levels. Each data level is assigned a colour and opacity value. The opacity value is used to give the user control over the weight of the data levels and to what limit a particular data level should be emphasised. The opacity value of each level ranges from 0 to 1. For full details on this transfer function, see the discussion of radiative transfer shaders in \citet{gooch:1995}, and the work by \citet{levoy:1988} and \citet{wittenbrink:1998}.

While this method gives more control over the volume rendering process and gives the user a better data insight, it has some disadvantages.
Firstly, it is harder to be utilized by inexperienced users, and secondly its rendering and compositing\footnote{Compositing is the process of combining two or more image components into a final image.} processes are more complicated.

\subsection{The frame composition processes}
\label{sct:VolumeRenderingandFramecompositingprocesses}

Within \citet{hassan:2011a,hassan:2012a} we discussed a direct send approach to implement distributed volume rendering of larger than memory datasets. While the framework introduced in these papers can still be extended to support other types of transfer functions, it might not produce the best frame rate. The main bottle-neck while implementing a more generic transfer function is the need for a specific compositing order, which limits the overlapping between computation and communication and delays the start of the composition processes until all the rendering processes are finished. Furthermore, in this case, the composition process is more sophisticated compared to the composition of MIP sub-frames. For these reasons, we introduce the new peer-to-peer communication processing mode.

\begin{figure}
\includegraphics[scale=0.27]{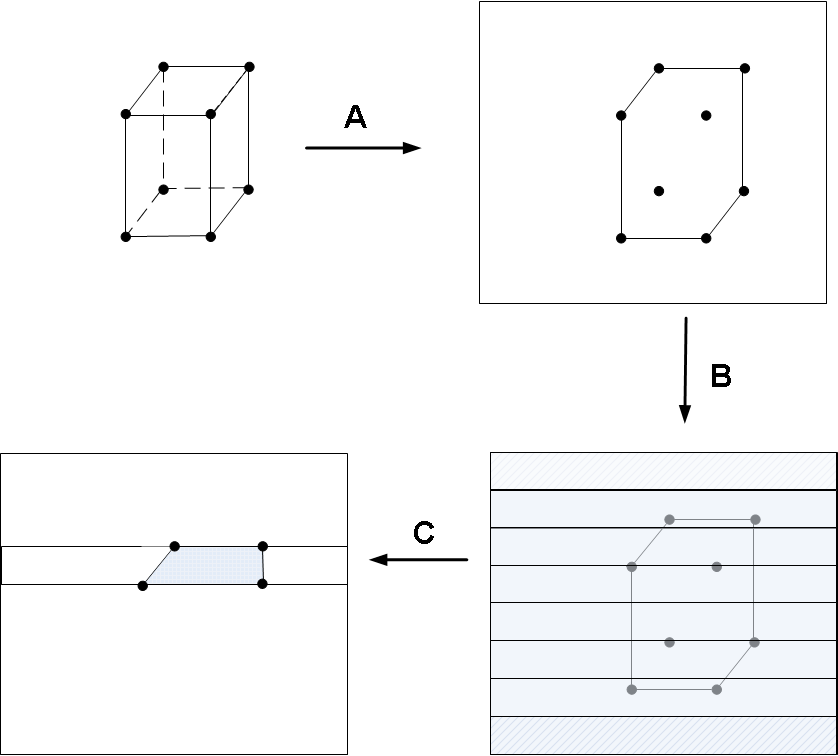}
\caption{A diagram showing the process of generating a rendering rectangle from a data sub-cube and the results of intersecting it with a single merging sub-rectangle. (A) The data cube is projected into the selected rendering viewport where a convex hull is computed to determine its bounding polygon. (B) After the rendering process is finalized, the rendering rectangle is intersected with all the merging sub-rectangles. (C) a sample intersection between the rendering polygon and a sample merging sub-rectangle.}
\label{fig:RenderingPolygon2} 
\end{figure}

\begin{figure}
\includegraphics[scale=0.2,viewport=0 0 1128 1004]{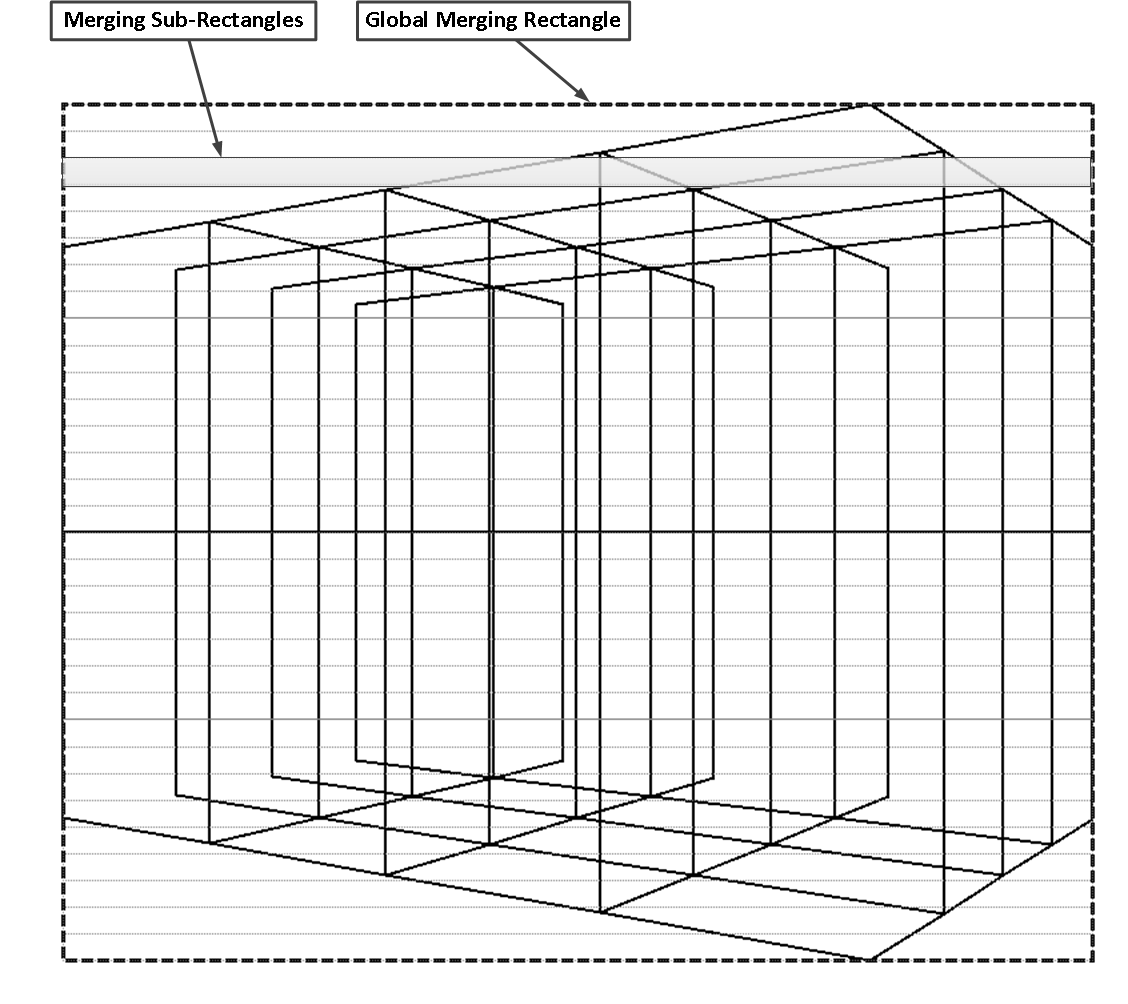}
\caption{An example of the rendering polygons generated from 32 sub-cubes and their global merging rectangle. The figure also shows how the global merging rectangle is partitioned into 32 disjoint sub-rectangles with an emphasis on one of them as an example. The highlighted sub-rectangle is used to show an example of the intersection between the rendering polygons and the merging sub-rectangle. }
\label{fig:MergingTerms} 
\end{figure}

Before we discuss the rendering and composition processes in more details, we need to clarify some terms that will be used within the description. 
\begin{itemize}
\item \textbf{Rendering Polygon:} as shown in step A of Figure \ref{fig:RenderingPolygon2}. The rendering polygon is a bounding convex polygon for the two-dimensional  region, on the output result, where the footprint of the current data cube is expected to contribute to the final visualization output. 
\item \textbf{Global Merging Rectangle:} as shown in Figure \ref{fig:MergingTerms}. It is the minimum bounding rectangle which bounds all the rendering polygons of the current frame.
\item \textbf{Merging Sub-Rectangle:}  as shown in Figure \ref{fig:MergingTerms}. It is a bordering rectangle that specifies a section, resulting from partitioning the area defined by the global merging rectangle into N disjoint regions. N is the number of rendering clients. The intersection of these sub-rectangles with the rendering polygons generates sub-rendering polygons as shown in step B and C in Figure \ref{fig:RenderingPolygon2}. 
\end{itemize}

From both technical and implementation perspectives, this work differs from the work presented in \citet{hassan:2011a,hassan:2012a} in the following points:
\begin{itemize}
\item The server is no longer required to do a lot of the processing, and is not required to run a GPU part to execute the composition processes. 
\item The data loading processes is no longer synchronized. Each process independently loads its own data portion as fast as the disk I/O allows, which speeds-up the data loading. 
\item The new implementation requires each processing client to know the data portions loaded by each of the other processing clients. So the data partitioning has been moved from the server to the processing clients. Each one of the processing clients executes the data partitioning part, and keeps the whole data map in its memory. 
\item  The data merging tasks require some sequential per-processing steps, which are executed by the CPU thread asynchronously with the rendering processes to reduce its overhead on the overall rendering time.
\item The framework is required to exchange both the colour and the opacity information of each pixel in the rendering sub-frames, which almost doubles the communication overhead.
\end{itemize}

\begin{figure}
\includegraphics[scale=0.4]{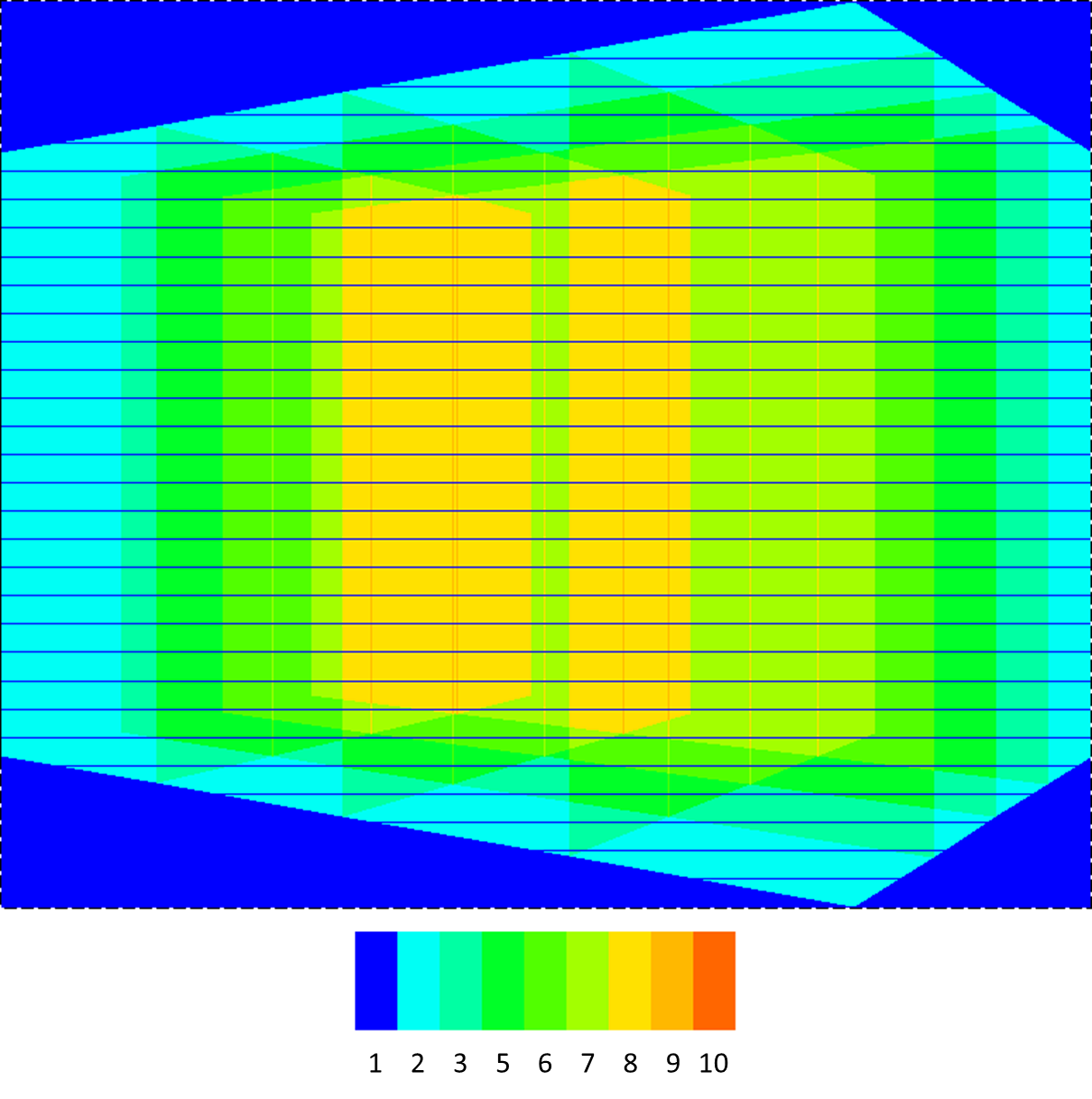}
\caption{A colour coded illustration of the rendering polygons shown in Figure \ref{fig:MergingTerms}. The colour presents the number of rendering polygons intersections per each pixel according to the colour map shown. The number of intersected rendering polygons varies from zero for the blue regions which are blank regions and narrow lines of red coded pixels where nine different rendering polygons intersect at the same pixel.}
\label{fig:ColorCodedMerge} 
\end{figure}

Within the rendering processes, more GPU functionalities were added to implement an additional two transfer functions. First, each processing client is still responsible for rendering its data on the selected viewport with the same rendering parameters. Then, these local results are composited to generate the final volume rendering output, noting that the composition operator may be required to be applied in either a front-to-back or in a back-to-front order on all the rendering sub-frames. Figure \ref{fig:ColorCodedMerge} shows a visual colour coded presentation of a sample output from the composition processes. The colours present the number of sub-frames that need to be merged to generate each data pixel on the output buffer within the global merging rectangle. See the shown colour-map for the colour to number of sub-frames mapping. 

The main idea behind the merging process is that each processing client will be responsible only for a portion of the final rendered frame which we refer to as the Merging Sub-Rectangle. Each processing client is responsible for two main geometric operations that define its communication patterns:
\begin{itemize}
\item Intersect its rendering polygon with all the rendering sub-rectangles to determine the recipients of its rendering results. The intersection is a set of convex polygons, which is packed into a linear buffer to be sent back to the other processing clients.
\item Intersect its assigned merging sub-rectangle with all the rendering polygons to know the expected number of rendering messages to be received and from which processing clients they are expected.
\end{itemize}

The communication pattern here is dynamic and based on the data partitioning and the rendering viewport. Each processing node will wait until all the needed merging buffers are received to start the merging process over the GPU and send the results back to the server. Regions with no rendering polygon intersections are excluded from the communication and the computation.
The results sending and receiving is performed asynchronously. So, the processing client is not required to wait for all other recipients of its results until it starts its own merging phase. 

\section{Quantitative visualization to accelerate or support decision making}
\label{sct:QuantitativeVisualization}
As discussed in \citet{hassan:2011b}, the lack of quantitative visualization techniques is one of the main limitations to the usage of 3D visualization tools in the day-to-day scientific activities of astronomers.
On the other hand, traditional two-dimensional tools(e.g. Miriad, Kvis, and CASA viewer) are widely used in radio astronomy, despite their lack of large dataset support. The reason behind this is their ability to mix data analysis tasks and visualization task, and produce publication-quality static output.

Within this work, we demonstrate how our distributed GPU-framework was enhanced to support different global (over the whole dataset) and local (over small user defined regions) data analysis and processing tasks. It is not our aim to implement all possible data analysis tasks. The functionalities presented within the next subsections are indicative of the data analysis and processing tasks that can take advantage of the GPU back-end. They also show how integration with the volume rendering output, facilitates enhanced data analysis. 

The functionalities selected in this prototype are:
\begin{itemize}
\item \textbf{Histogram:} as an example of the data analysis processes which require visiting each data point in the underlying dataset only once in addition to exchange non-constant amount of data between different processing nodes. 
The histogram is a concise and useful summary of the noise, signal and artifact components of an image, and accordingly is an exceedingly common operation in image analysis, quality control, and noise estimation. 
\item \textbf{Global Mean and Standard Deviation:} as an example of the data analysis processing which requires summarizing the whole dataset into a single data value.
\item \textbf{Global Median:} as an example of the data analysis tasks that need multiple iterations to converge to the correct solution.
\item \textbf{Sigma-Clipping Transfer Function:} as an example of data analysis tasks that need local information and integration of the volume rendering output. It can be used to estimate the best data threshold level to be utilized as an input to a later source extraction process.
\item \textbf{3D Spectrum Tool:} as an example of quantitative data interaction with the displayed volume rendering output and how to query (i.e select a portion of the dataset based on a search parameters) the data based on this interaction. This tool demonstrates the framework ability to perform rapid profile extraction, which is a common data analysis operation. 

\end{itemize}

The main challenges that we have addressed are:
\begin{itemize}
\item The selection of an appropriate algorithm, within each task, that is suitable to be implemented on both distributed memory and shared-memory architecture.
\item The dataset portions loaded over different GPUs contain some overlaps and repetitions (due to volume rendering requirements) that need to be excluded correctly to produce accurate statistical information.  
\item In some cases it is necessary to make a special handling for the NULL (undefined) data points to produce a correct statistical output, which is rarely addressed in ready made implementations. 
\item Addressing large dataset puts a restriction on the kind of algorithms that can be implemented in a memory limited environment.  
\end{itemize}
  
In the next sections, we present how we address these different issues for each of the prototyped functionalities.

\subsection{Histogram}

\begin{figure}
\includegraphics[scale=0.25,viewport=0 0 920 583]{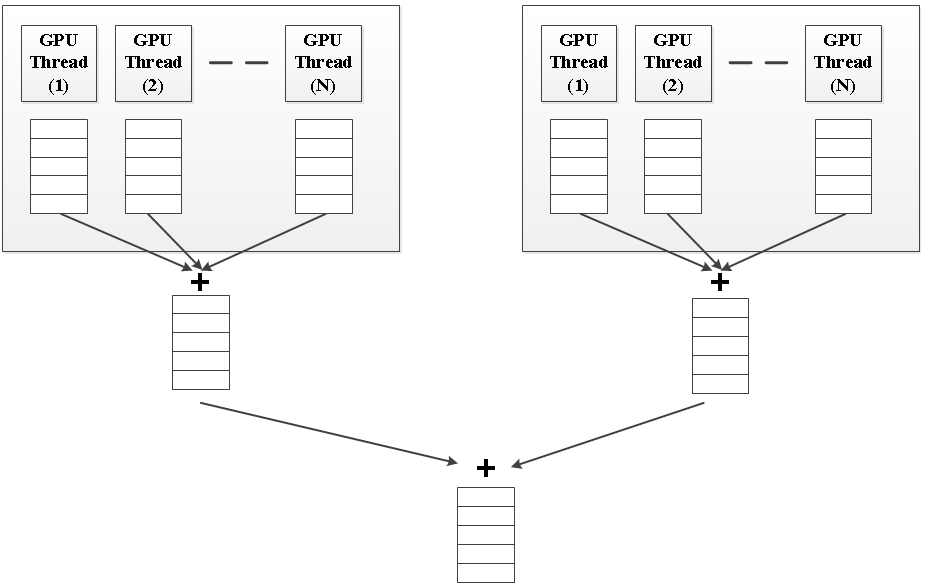}
\caption{Schematic diagram for the distributed process of calculating global histogram combined with a simple pseudo code to describe the algorithm in its sequential form. }
\label{fig:Histogram} 
\end{figure}

Figure \ref{fig:Histogram} shows a simplified schematic diagram for the distributed process of calculating a global histogram, combined with a simple pseudo code to describe the algorithm in its sequential form. 

This algorithm is implemented using the client-server processing mode discussed in Section \ref{sct:DistributedGPUframework}. 
Each processing node calculates a local histogram of its data, while the repeated data portions are calculated only once. The data is then sent to the server to merge local histograms to a global histogram. Within the single processing node level, the nodes' data is partitioned into smaller data portions and the problem is further distributed over GPU cores. If the data minimum and maximum are not known in advance, they are calculated while the dataset is loading. Because the result's composition operator (summation) is associative and commutative, any arbitrary data composition order is valid. Furthermore, the amount of data to be sent from the nodes to the server is relatively small, which make the communication overhead in this case negligible. As we will show in Section \ref{sct:Results}, we can calculate a histogram for a 500 GB image in four seconds. 

\subsection{Global Mean and Standard Deviation}

The process of calculating the global data mean and standard deviation is similar to the process of calculating the data histogram. It requires visiting each data value once in order to compute the local summation and summation of squares and count of the non-null data values within each sub-cube. 
The distributed execution of these computation is similar to the histogram algorithm. 

The final evaluation of these computations for the global dataset is done on the server, which is an $O(1)$ operation. The data composition operator here is the summation operator which is associative and commutative so the client-server computing mode is utilized.

Accumulating large amount of data points might lead to numerical overflow. To avoid this, the framework uses mixed precision approach. It partitions each sub-cube into 2D slices.  The data values of each sub-cube slice are summed into a single precision variable. The accumulation of each slice's summation is preformed into a double precision variable which is sent to the server for the final data accumulation.

\subsection{Global Median}

The median computation over distributed GPU infrastructure is a challenging problem because of the limitations imposed by the data size and the GPU architecture. We can summarize these limitations into:

\begin{itemize}
\item The sort-based median calculation, i.e. sort and select the middle element, is prohibitively expensive  over a distributed framework because of the communication overhead of the sorting part and the lack of sufficient memory space. 
\item Partition-based median calculations \citep{Hoare:1961}, require having a sorted replicant (or even a sub-set) of the current data array in local memory. The needed memory capacity for storing such a replicant might not be available. In addition, reloading the dataset will add a big I/O cost. 
\item To support fast volume rendering, we stored the sub-data cubes over a special GPU memory called texture-memory, which is read-only from the GPU cores perspective. So, data editing over GPUs is not allowed. 
\end{itemize}  

It might be possible to relax some of these limitations by utilizing the CPU memory in addition to the GPU memory (assuming that the user has access to enough CPU memory to hold a dataset replica). However, we chose not to relax any of these conditions for prototyping and practicality purposes. 

The median algorithm utilized in this work was introduced by Torben Mogensen\footnote{\url{http://ndevilla.free.fr/median/median/index.html}}. The algorithm is described by pseudocode in Appendix \ref{App:TorbenMethod}. Although it is the slowest sequential median finding algorithm, it has a couple of features that make it interesting for a GPU and distributed memory implementation:
\begin{itemize}
\item It is an in-place algorithm and does not require data to be edited or sorted;
\item It can be easily parallelized over both shared and distributed memory architectures with a negligible communication cost; 
\item The result composition is performed by an associative and commutative operator which enables using the client-server processing mode; and
\item It can be easily modified to address the median absolute deviation (MAD).
\end{itemize}

This algorithm converts the median problem into a problem of visiting each data point and classifying it into one of three categories: equal, greater, or lower than the current median guess. This kind of computation is a good fit to the GPU architecture and requires only five data values to be exchanged for each iteration. These data values represent: the count of voxels less than, greater than, and equal to the current median guess, the value of maximum voxel less than the current guess, and the value of the minimum voxel greater than the current guess within each sub-cube. The only drawback in this algorithm is that it is an iterative algorithm and the load of computation in each iteration is constant $O (N)$. The number of iterations depends on the data values; it is estimated to be $O(log N)$. Table \ref{tblMedian} shows a sample run for Torben's Method. The dataset used is a high-resolution 21-cm data cube of the Large Magellanic Cloud (11.1 deg x 12.4 deg by 196 km/s) constructed from the combined Parkes and Australia Telescope Compact Array (ATCA) survey \citep{kim:2003}.  The dataset comprise 534,397,200 data points, a data minimum of -1.0449826717, and a data maximum of 0.8189788461 . The algorithm took 31 iterations to reach the exact data median. As it is clear from the guess column in the data, if a median approximation is accepted, the number of iterations could have been reduced to 20 with an error of order $10^{-6}$.  
  
We used the same two-stage  composition described in Figure \ref{fig:Histogram} to combine the results of each iteration. The iteration main control is managed by the server. Within each iteration, the server passes the median guess and waits for five variables, which are the number of data points greater than, less than, and equal to the current guess, and two floating point values representing the maximum value of the data points less than the current guess and the minimum value of the data points greater than the current guess. These values are used to determine the next guess.

\begin{table*}

\caption{A sample run for Torben's Method  over  534,397,200 data points with a data minimum -1.0449826717 and a data maximum 0.8189788461. The data median is 0.0036070324 which is correctly calculated after 31 iterations. The dataset used is a high-resolution 21-cm data cube of the Large Magellanic Cloud (11.1 deg x 12.4 deg by 196 km/s) constructed from the combined Parkes and Australia Telescope Compact Array (ATCA) survey \citep{kim:2003}.\label{tblMedian}}
\begin{tabular}{|c|c|c|c|c|c|c|}
\hline
Iteration & Guess & Minimum & Maximum & Less Count & Greater Count & Equal Count \\ \hline
1&-0.1130019128 &-1.0449826717&   0.8189788461 &   210274&534186926&0 \\
2&0.3529884815  &-0.1130018905&   0.8189788461 &   533594229&802970&1\\
3&0.1199932843 & -0.1130018905&   0.3529884517 &   525168671&9228528&1 \\
4&0.0034956932 & -0.1130018905&   0.1199932769 &   265962031&268435160&9\\
5&0.0617444851 & 0.0034956937 &   0.1199932769 &   509259427&25137770&3\\
10&0.0044058301& 0.0034956937 &   0.0053159669 &   276040288&258356903&9\\
20&0.0036067935& 0.0036059052 &   0.0036076820 &   267195959&267201241&0\\
30&0.0036070314& 0.0036070300 &   0.0036070328 &   267198590&267198608&2\\
31&\textbf{0.0036070324}& 0.0036070317 &   0.0036070328 &   267198599&267198597&4\\ \hline
\end{tabular}
\end{table*}
\subsection{3D Spectrum Tool}
\label{subsect:Spectrumtool}
\begin{figure*}
\includegraphics[scale=0.28,viewport=0 0 1750 705]{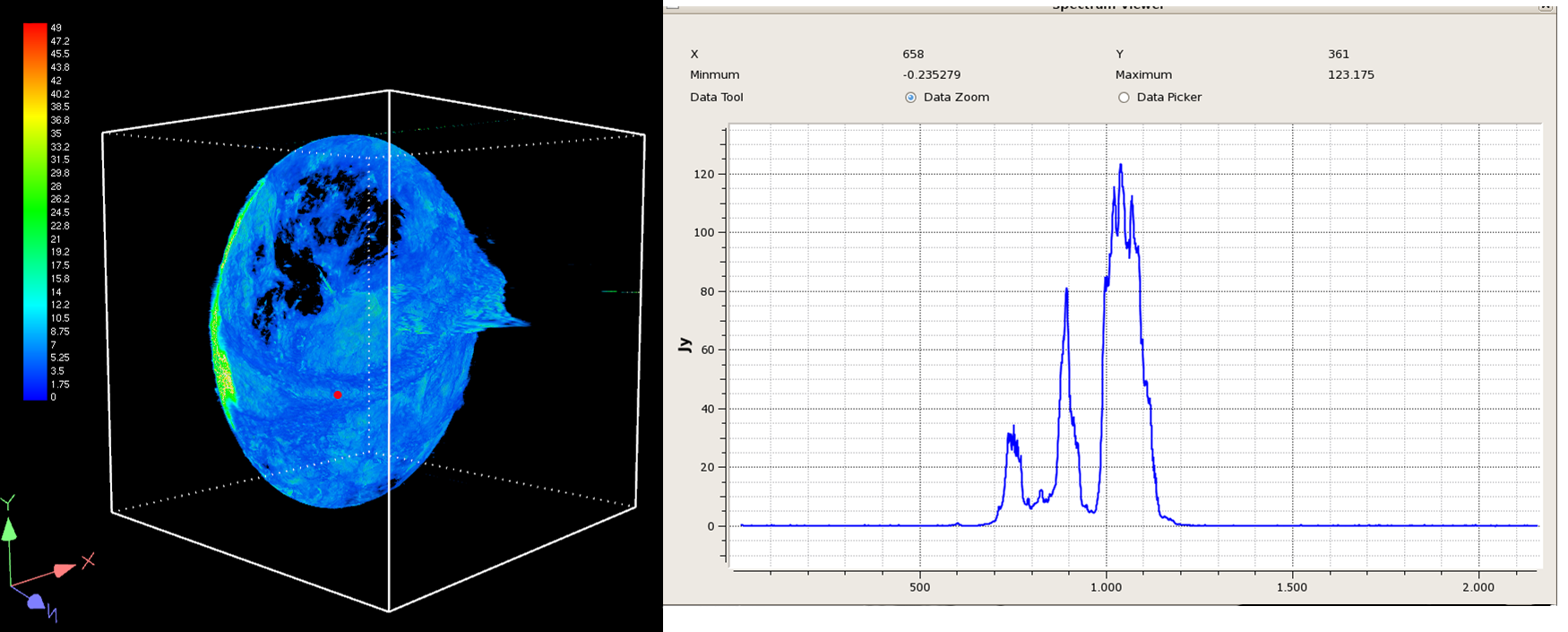}
\caption{Sample output of the 3D spectrum tool described in Section \ref{subsect:Spectrumtool}. On the left, the figure shows a volume rendering output for the GASS cube with a red dot indicating the place that was picked by the user. On the right, the spectrum is generated for a line-of-sight ray starting from this position and going through the data cube. The tool features a zooming function and a point picker to show the exact data value at a specific location.   }
\label{fig:3DSpectViewer} 
\end{figure*}

\begin{figure}
\includegraphics[scale=0.5,viewport=0 0 343 522]{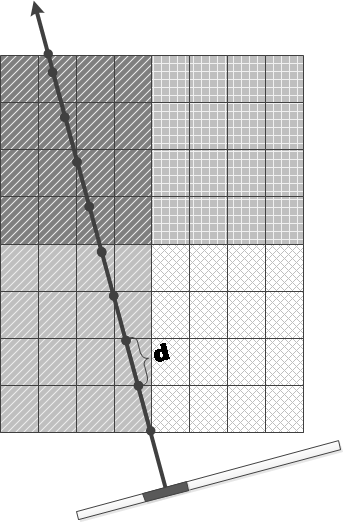}
\caption{Two-dimensional illustration of the processes of evaluating the sampled data points on an arbitrary ray with a constant sampling step. The presented tool uses the same technique in 3D with trilinear interpolation to determine the points' values. Within the shown illustration, the data grid is partitioned into four different sub-grids with the ray passing through two of them. For simplicity, the overlapping between different sub-grids has been omitted.}
\label{fig:3DSpect} 
\end{figure}

This functionality presents an integration between the visualization output and the ability to query the data stored in multiple processing clients based on the user selection. This tool is similar to the ``profile window'' functionality in kvis\footnote{\url{http://www.atnf.csiro.au/computing/software/karma/user-manual/node3.html#sectionprofiles}}. Instead of supporting a data profile in only one of the axis directions (X,Y, or Z), this tool provides a data profile in any arbitrary 3D data direction. Figure \ref{fig:3DSpectViewer} shows a sample output of this tool with the Galactic All-Sky Survey data (see Table \ref{tab:ListOfDatasets}). The user input for this tool is a specific screen position, selected by the mouse interaction in the GUI client. The chosen mouse position is converted to data coordinates, and the system computes a data profile for a ``line of sight'' with the selected pixels as the origin and with a direction perpendicular to the data cube. Figure \ref{fig:3DSpect} shows a simplified example of this process in 2D. The system use trilinear interpolation to interpolate the data values of the generated profile from the original data points. 

\subsection{The Sigma-clipping transfer function}
\label{sct:SigmaClippingTF}
Source finding is an important data processing step for radio astronomy spectral cubes. 
It is concerned mainly with identifying interesting data regions for further user attention.
Through the literature, we can identify two main source finding methodologies:
\begin{itemize}
\item \textit{Template matching approach} where an automated process searches for a specific 2D or 3D pattern within the data cube [e.g the work done by \citet{saintonge:2007} for the Arecibo Legacy Fast ALFA Survey]. 
\item \textit{Thresholding (Sigma-Clipping) techniques} where a data threshold is defined based on the estimated noise properties, and this threshold is utilized to identify important data regions with a certain probability based on how many times their data values are larger than the estimated noise levels [e.g. \textsc{duchamp}; \citet{whiting:2012}]. 
\end{itemize}

The Sigma-Clipping technique is more widely utilized through tools such as \textsc{imsad}\footnote{\url{http://www.atnf.csiro.au/computing/software/miriad/doc/imsad.html}} and \textsc{duchamp}. These tools require the user to estimate a couple of parameters that control the data smoothing and the thresholding processes. A detailed review of the current available source finding techniques is outside the scope of this paper. We forward the reader to \citet{koribalski2012overview} for further details.
We will restrict our attention to the Sigma-Clipping technique and how visualization can support it as a preprocessing step.

Users often face a challenge in determining the thresholding parameters, which are highly dependent on the dataset characteristics. We are arguing with this work that qualitative and quantitative visualization can facilitate and support the process of selecting such parameters. 

The transfer functions discussed in Section \ref{sct:TransferFunction} can be utilized to apply a global data threshold given an exact data value (in terms of the data measurement units). The previously discussed statistical data computation can be used to support user decisions on such a data threshold (e.g mean and standard deviation). Additionally, the generic transfer function can be utilized to display the data based on a user-controlled weighting. 
This approach can enhance source finder processes by speeding up the selection of an appropriate data threshold parameter. 
By interactively modifying the data threshold, the user can see what the source finder will see: the amount of noise remaining, and the number of data features that have disappeared. 
Additionally, the same technique can be used to help the user to eliminate regions with small-scale  artefacts (e.g. radio frequency interference) or to apply a special pre-processing handler to the large-scale  artefacts (e.g. removing ripples in the baseline). 
Furthermore, it is easy to implement some data smoothing filters (not currently implemented within our prototype) to help the user to see the effect of these operations. 

\begin{figure}
\includegraphics[scale=0.25,viewport=0 0 1002 379]{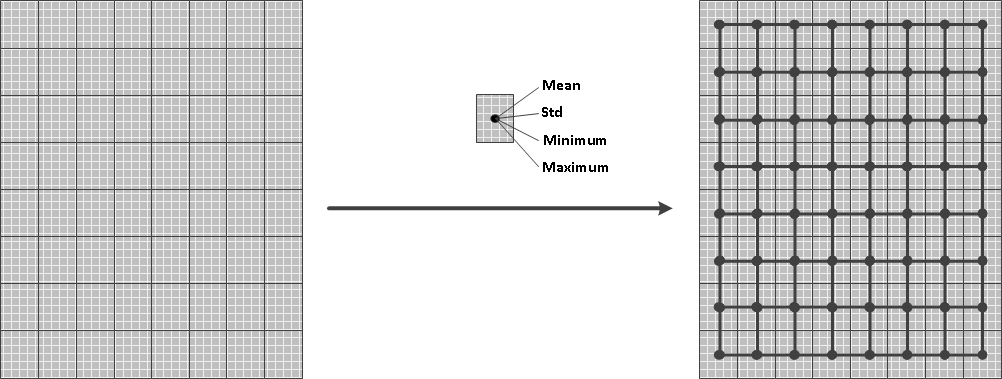}
\caption{Two-dimensional illustration of the mini-map concept. Each newly defined point in the mini-map (big black dot) summarize the properties of a larger data region (defined by the bounding rectangles). Two interpolation operations are required to compute the data point colour and opacity. The first one is done over the original grid and the second one is done on the mapped point coordinate to the mini-map.}
\label{fig:Minimap} 
\end{figure}

The Sigma-Clipping transfer function is more concerned with identifying a local data threshold rather than the global data threshold, which can be achieved with the other two transfer functions. 
This methods start by defining what we call a statistical mini-map, which is a scaled-down version of the data cube (per each compute node). Here, each data point represents the data statistics of a bigger region in the real dataset. 
The size of this data map is controlled by the user. 
The calculated summary of the data cube is then utilized to define a local data threshold.  
The mini-map concept is illustrated in 2D in Figure \ref{fig:Minimap}. For each sampled data point over the traced rays, two trilinear interpolation operations are required. The first one is performed to get the actual data value at this point using the original data grid. The second is performed to get the data properties at this point using the mini-map. The point's data value is then scaled using its region's data properties to generate the look-up value used to calculate the colour and the opacity at this point.

\begin{figure}
\includegraphics[scale=0.3,viewport=0 0 850 408]{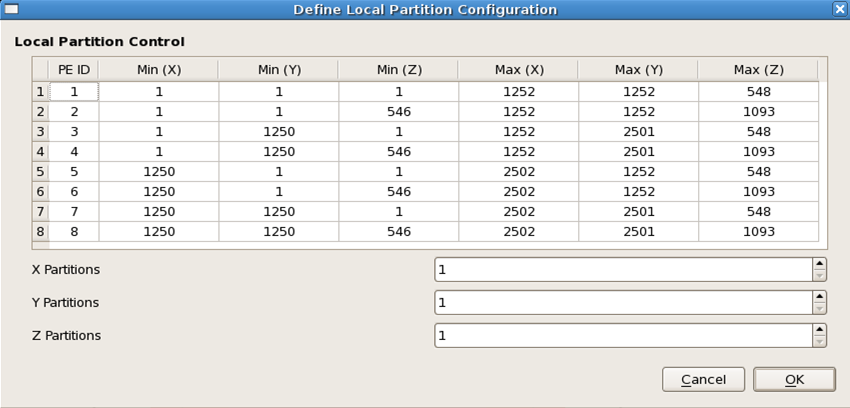}
\caption{The GUI used by the user to define the local data partitions. The interface gives the user information about the number of processing nodes, and the size of the data sub-cubes loaded on each of them. The user is required to submit the number of partitions required for each of the main data dimensions. The system automatically determines the size of the mini-map and then calculates the needed statistical summary.  }
\label{fig:LocalPartitionInterface} 
\end{figure}

Figure \ref{fig:LocalPartitionInterface} shows the interface used to control the data partitioning and the data map generation. 
The information given to the user within this interface is the number of processing clients available and the data portion loaded in each of these clients. 
The user can specify how many data partitions are needed in each of the basic coordinates. 
This information is used to pre-compute the data map before giving the user the ability to interact further via the sigma-clipping interface.

\begin{figure}
\includegraphics[scale=0.5]{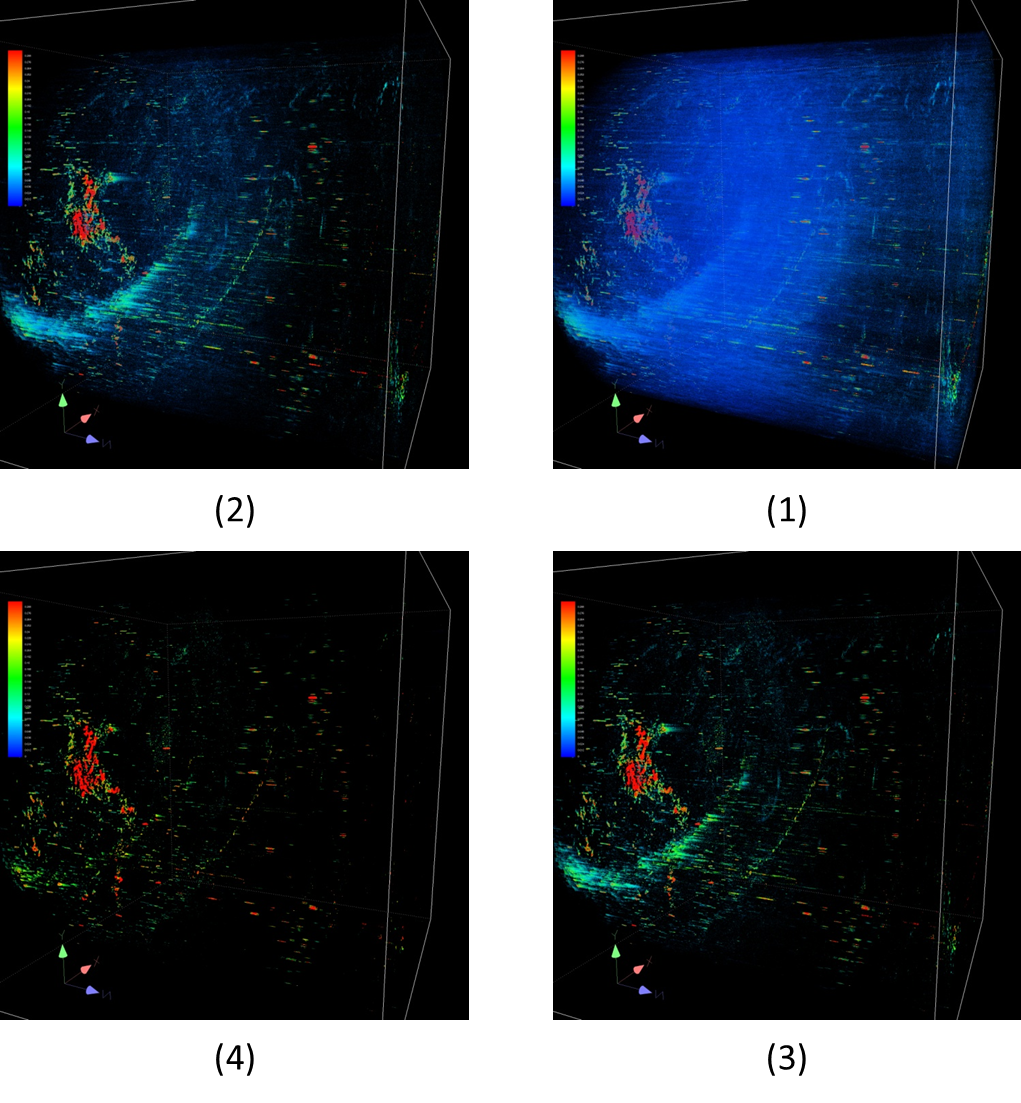}
\caption{Sample outputs for the Sigma-Clipping transfer function using the HIPASS dataset (see Table \ref{tab:ListOfDatasets}). The figure shows  the output of the rendering process with different sigma clipping thresholds: 2$\sigma$,3$\sigma$, 4$\sigma$, and 7$\sigma$. The local partitions are defined to be 30$\times$30$\times$30 pixel size. The Milky Way channels were removed to allow better illustration.    }
\label{fig:SigmaClippingFig} 
\end{figure}

Figure \ref{fig:SigmaClippingFig} shows sample outputs for the Sigma-Clipping transfer function volume rendering using the HIPASS dataset (see Table \ref{tab:ListOfDatasets}). The figure shows the output of the volume rendering with 4 different clipping thresholds: 2$\sigma$,3$\sigma$, 4$\sigma$, and 7$\sigma$. The local partitions are defined to be 30$\times$30$\times$30 pixel. For a better illustration output the Milky Way channels were removed. 

The main drawback of this technique is its need for a better user understanding of visualization concepts.  
On the other hand, this technique gives the user a fine control over the volume rendering process and involves utilizing both the local and the global data characteristics.
 
\section{Results and Discussion}
\label{sct:Results}

In order to demonstrate that the framework is scalable enough to work over a large number of nodes and GPUs, and can handle tera-scale datasets, we conducted a series of tests and performance benchmarks, which are described and discussed within this section.

\subsection{Benchmarking and Performance analysis}
\begin{figure}
\includegraphics[scale=0.3]{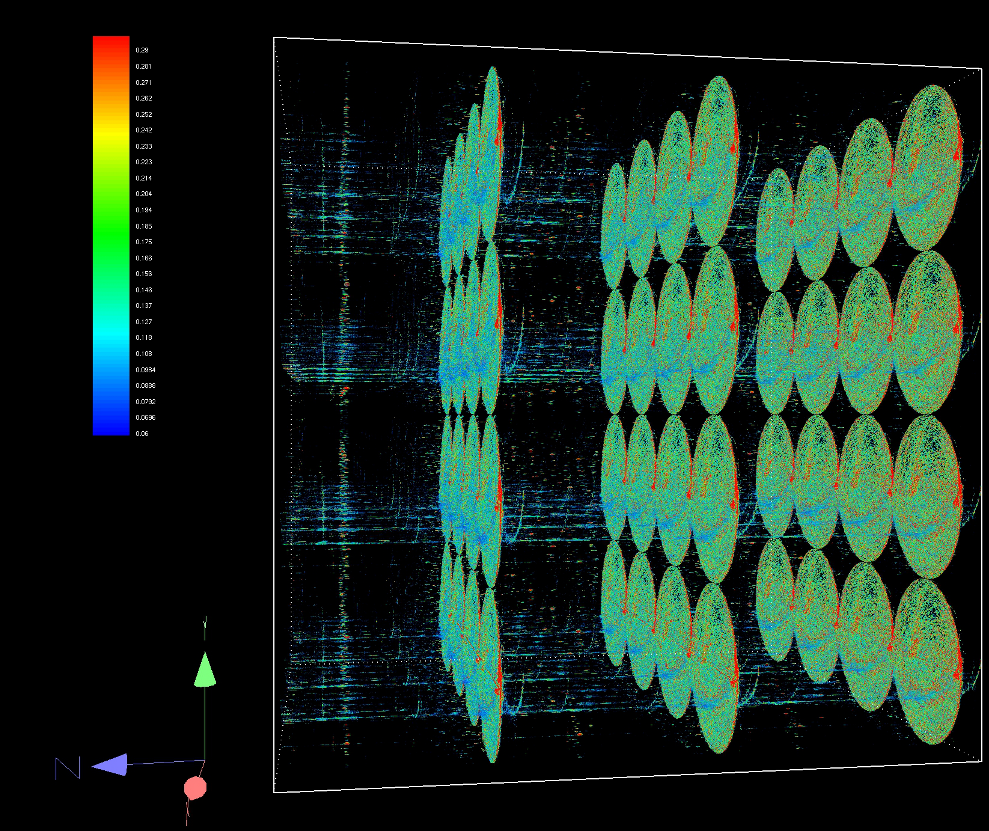}
\caption{ A sample rendering output for the 48X HIPASS dataset performed with the framework described in this paper over a 96 GPUs with custom transfer function with linear increasing opacity.}
\label{fig:HIPASSSampleOutput} 
\end{figure}

The framework performance analyses and benchmarking were conducted with the Swinburne gSTAR hybrid CPU/GPU supercomputer. These analyses were conduced using  datasets built from the HI Parkes All Sky Survey (HIPASS) data cube \citep{meyer2004hipass} by replicating it multiple times in each of the basis directions (X, Y, and Z). The original HIPASS data cube was generated by Russell Jurek (ATNF), from 387 HIPASS sub-cubes\footnote{\url{http://www.atnf.csiro.au/research/multibeam/release/}}, with a dimension of 1721 x 1721 x 1024 and size of 11.3 GB. Table \ref{tab:ListOfDatasets} lists the dimensions, and the data size of each of the scaled cubes. The maximum dataset size involved in these tests is a 542 GB file, which represents 48 times the HIPASS data cube. Figure \ref{fig:HIPASSSampleOutput} shows a sample volume rendering output of this cube with a generic transfer function.

\begin{table*}
\caption{Sample datasets used to evaluate the performance of our framework.}
	\label{tab:ListOfDatasets}
	\centering
		\begin{tabular}{ccp{5cm}c}
		\hline
			Dataset Name & Dimensions (Data Points) & Source / Credits & File Size\\ \hline	
			HIPASS Cube  & 1721 x 1721 x 1024 & HIPASS Southern Sky, data courtesy Russell Jurek/HIPASS team  & 11.3 GB \\ 			
			8X HIPASS Cube  & 3442 x 3442 x 2048 & Replicated HIPASS Southern Sky Cube, (2$\times$2$\times$2)  & 90.4 GB \\ 
			27X HIPASS Cube  & 5163 x 5163 x 3072 & Replicated HIPASS Southern Sky Cube, (3$\times$3$\times$3)  & 305.1 GB \\ 	
			48X HIPASS Cube  & 6884 x 6884 x 3072 & Replicated HIPASS Southern Sky Cube, (4$\times$4$\times$3) & 542.33 GB \\ \hline	
		\end{tabular}	
\end{table*}

The configuration of the gSTAR nodes used in these timing analyses is shown in Table \ref{tab:NodesConfig}. The nodes are connected using a QLogic InfiniBand QDR\footnote{\url{http://www.qlogic.com/Resources/Documents/DataSheets/Switches/12300_Datasheet.pdf}} switch with a theoretical bandwidth of 40 Gbps. Furthermore, the system provides 1.7 petabytes of usable disk space served by a Lustre file system\footnote{\url{http://wiki.lustre.org/index.php/Main_Page}}. A maximum of 48 nodes, each with two GPUs, were allocated for this benchmark with a total theoretical GPU memory of 588 GB\footnote{A small portion of it is not accessible to the user programs.}.

\begin{table*}
\caption{The different configuration of gSTAR nodes used with the performance analyses and benchmarking }
	\label{tab:NodesConfig}
	\centering
		\begin{tabular}{ccccc}
		\hline
			Number of Nodes & Total Number of GPUs & GPU Model & GPU memory & Number of Cores\\ \hline	
			45 & 90 & NVIDIA Tesla C2070  & 6GB & 448\\ 
			3 & 6 & NVIDIA Tesla C2090  & 6GB & 512\\ \hline
		\end{tabular}	
\end{table*}

Due to the Lustre distributed file system, our framework is capable of loading the 48X HIPASS dataset (542 GB) into the GPU memory in around 540 seconds $\simeq$ 9 minutes. Table \ref{tab:AnalysisSummary} summarizes the data loading time for the datasets used in these timing tests. Datasets 48X HIPASS, 27X HIPASS, and HIPASS were distributed over 113 file strips\footnote{The number of object storage targets used by the Lustre file system to store the data file. See \url{http://wiki.lustre.org/index.php/Configuring_Lustre_File_Striping} for more details. This number is an indication of the distribution level of the file's data blocks.} (the maximum allowed by gSTAR's current Lustre configuration). To show the impact of using a distributed file system, the 8X HIPASS data is stored in a single object storage target. Due to the file distribution configuration, for datasets HIPASS and 27X HIPASS, the more nodes used the less loading time required. For the 8X HIPASS dataset the more  nodes used, the longer the data loading time required. While it is still possible to load the 8X HIPASS using a sequential accessing mechanism, our trials to scale this for the 27X HIPASS and 48X HIPASS tended to cause a file I/O deadlock.
 
Table \ref{tab:AnalysisSummary} also shows the execution time for the different quantitative data analysis operations described in Section \ref{sct:QuantitativeVisualization} for different datasets and number of GPUs. These data analysis operations include calculating median, mean and standard deviation, and histogram. The table also shows average data size loaded per each GPU, which is directly proportional with the processing load required from each GPU. 
The performance measurements and timing results shown in Table \ref{tab:AnalysisSummary} are not constant numbers. Small variations in these numbers are expected due to the variable communication overhead between the different framework components.

\begin{table*}
\caption{Summary of the data loading time, data load per GPU for the datasets used in this benchmark (see Table \ref{tab:NodesConfig}) with different number of GPUs. Also, the table summarizes the execution time of different data analysis tasks including: median (exact value), mean/standard deviation, and histogram.}
\label{tab:AnalysisSummary}
\centering
\begin{tabular}{cccccccc}
\hline
Dataset & Lustre File Strips & GPUs & File Loading (s) & Median (s) & Mean/Std (s) & Histogram (s) & Load/GPU (GB)\\ \hline	
48X HIPASS & 113&96 & 546 & 44.794 & 1.745 & 4.013 & 5.65\\ 
27X HIPASS & 113 &64 & 523 & 38.9&1.3&4&4.77\\ 
27X HIPASS & 113 &96 & 323 & 22.75&1.26&3.98&3.177 \\ 
8X HIPASS & 1 & 32 & 384 & 23.5&0.65&1.89&2.82\\ 
8X HIPASS & 1 & 64 & 484 & 11.5&0.53&1.64&1.41\\ 
8X HIPASS & 1 & 96 & 502 & 7.87&0.51&1.63&0.94\\ 
HIPASS & 113& 8&36& 12.571&	0.49&	1.37&	1.41\\ 		
HIPASS & 113&  16&21&	6.325&	0.25&	0.78&	0.71\\ 
HIPASS &113 &32&16&3.453&	0.13&	0.41&	0.35\\ 
HIPASS &113&64&15&1.658&	0.15&	0.39&	0.18\\ 
HIPASS &113&96&10&1.946&	0.13&	0.39&	0.12\\ \hline			
\end{tabular}	
\end{table*}

\begin{figure*}
\includegraphics[scale=0.15]{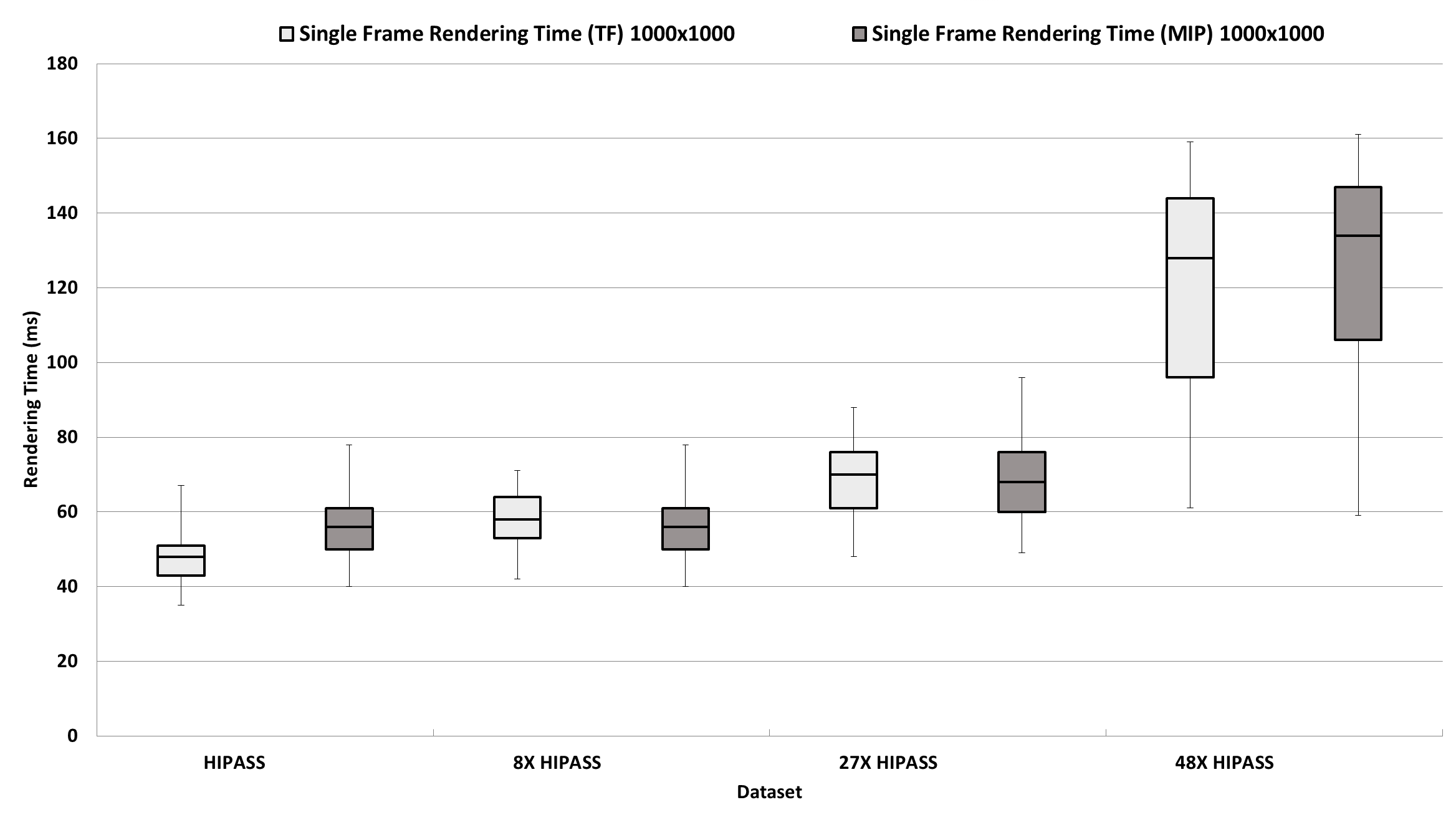}
\caption{Show minimum,25th percentile, median, 75th percentile, and maximum rendering time (in millisecond) for all datasets with 96 GPUs, and 1000 x 1000 output pixel with both generic and maximum intensity projection transfer functions.}
\label{fig:HIPASSRenderingTime_All96_TF} 
\end{figure*}

Figures \ref{fig:HIPASSRenderingTime_All96_TF} shows the minimum, 25th percentile, median, 75th percentile, and maximum rendering time (in millisecond) for all the datasets used with 96 GPUs and 1000x1000 pixel output resolution.  

\begin{figure}
\includegraphics[scale=0.16]{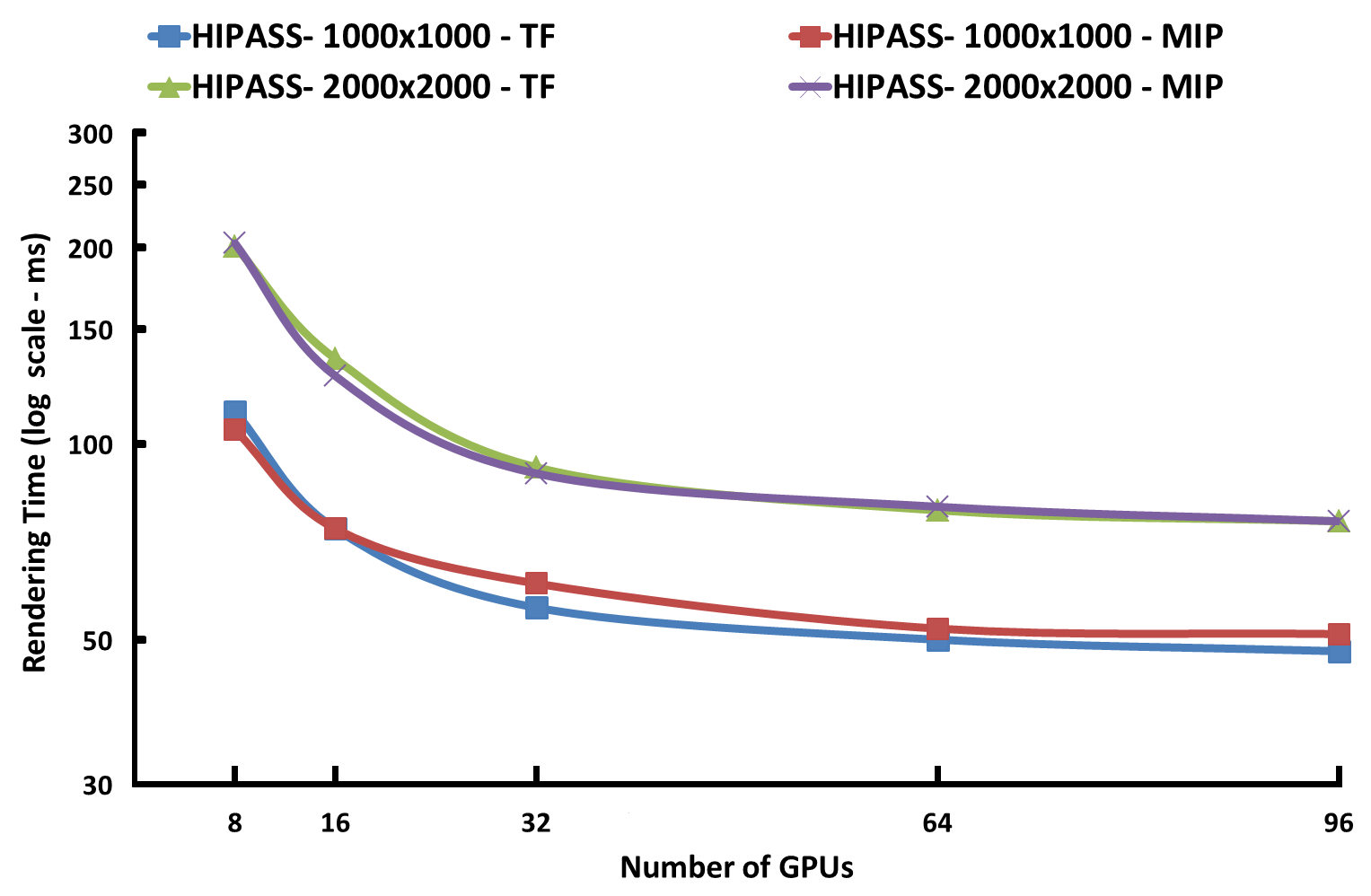}
\caption{A log-linear plot of the 75th percentile Rendering time (in millisecond) for the HIPASS dataset with different number of GPUs (presented on the X-axis), different output resolution (1000 x 1000 and 2000 x 2000), and different transfer functions (maximum intensity projections and generic transfer function).  }
\label{fig:HIPASSRenderingTime} 
\end{figure}

The data points in Figures \ref{fig:HIPASSRenderingTime_All96_TF} and \ref{fig:HIPASSRenderingTime} are computed in the following steps:
\begin{enumerate}
\item[1.] An automated script is used within the GUI interface to render 180 frames for each of the data cubes with the data cubes rotated around the Y-axis with a step angle of $2^{\circ}$. 
\item[2.] This process is repeated 5 times, to provide an average for each view angle. 
\item[3.] The average rendering time of the different angles is sorted and the 75th percentile (in milliseconds) is used as an indication of the rendering time as in Figure \ref{fig:HIPASSRenderingTime}. 
\end{enumerate}

\begin{figure}
\includegraphics[scale=0.35]{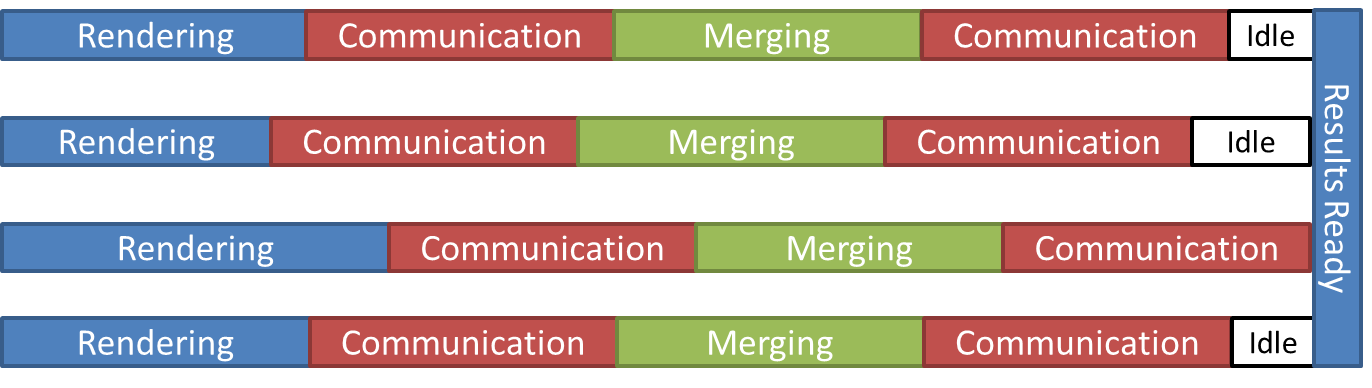}
\caption{An illustration of the execution time for each of the basic rendering sub-operation and how they affect the final rendering time. }
\label{fig:RenderingCalc} 
\end{figure}

As discussed in \citet{hassan:2012a} and \citet{hassan:2011a}, the rendering time of different frames varies based on the orientation angle of the dataset, which affects directly the ray lengths and the number of samples required per each ray to determine the final colour and opacity. Figure \ref{fig:RenderingCalc} illustrates the main components of the rendering process and the effect of their execution time on the final rendering time and hence the final frame rate of the rendering process. Most of these component's execution times are not constant, even for the same rendering configuration, due to changes in the cube orientation, transfer function selection, and output resolution. Furthermore, the communication between different components in the system is affected by external parameters (e.g. other jobs running on the system and their usage of the communication infrastructure, the order in which the communication happens) which cannot be easily controlled.

Figures \ref{fig:HIPASSRenderingTime} shows the rendering time (as the 75th percentile - in milliseconds) for the HIPASS data cube with a different number of GPUs (from 8 to 96). The HIPASS dataset is used to show the effect of increasing the number of processing nodes while keeping the workload constant. The HIPASS dataset is the only dataset, within our test datasets, that can fit within the memory of 8 GPUs (maximum of 48 GB). The figure shows the rendering time for the maximum intensity projection and the generic transfer function with two different output resolution: 1000 x 1000 and 2000 x 2000 pixel.

In general, the amount of communications required to synchronize between different parts in the framework increases with the number of processing nodes and GPUs. On the other hand, the workload assigned to each GPU is decreased with the increase of processing nodes and GPU number, which is shown in the final column of Table \ref{tab:AnalysisSummary}.
  
The main trend in Figures \ref{fig:HIPASSRenderingTime} is that the execution time decreased with the increase of the number of nodes even with the increased communication overhead. This happens mainly because the actual local rendering process (ray-casting execution on the GPUs) dominates any other components. The amount of rendering time reduction is relatively higher when moving from 8 to 16 GPUs but starts to decrease fast with the further increase of GPU counts until it reaches zero as shown between 64 and 96 GPUs. At this level, the amount of data allocated for each GPU is lower than 0.2 GB with the communication overhead starting to dominate the rendering time.

\subsection{Discussion}

We have designed, implemented and demonstrated a data analysis and visualisation framework that can:
\begin{itemize}
\item \textbf{Handle data larger than a single machine memory in near realtime.} We demonstrate this with different scales of datasets ranging from 12 GB to over 0.5 TB with a near real-time performance for most of the shown functionalities.
\item \textbf{Launch autonomous distributed analysis and visualization jobs.} This was achieved by using the remote data analysis and visualization concept. Within our implementation, the GUI client is separated from the server components where the actual processing happens. With such separation, the user's physical location is no longer a barrier for accessing large scale distributed infrastructures. Another benefit of this separation is that it hides the complexity of the back-end heterogeneous architecture from the user.
\item \textbf{Employ computational accelerators.} Within this framework, we demonstrate the benefit of using GPUs as the main processing element. The presented framework managed to combine the processing power of both GPUs and CPUs to provide the user with real-time or near real-time performance for different data analysis and processing tasks on a large scale datasets. 
\item \textbf{Minimize data movement, and deal effectively with distributed and remote data-storage facilities.} This is another benefit of using the remote data analysis and visualization concept. The remote client needs only a small portion of the dataset metadata which can be sent on demand using a moderate speed connection. So, the data is no longer required to be located on the user's desktop computer or in a local data-storage facility. 
We expect this to significantly minimize the need for data movement, it will speed-up the data analysis and visualization process, and minimize the cost and technological barrier for analysing large scale datasets.  
Moreover, within this work we demonstrates how using a distributed file system (Lustre) is effective in minimizing the data loading time (around 9 minutes for 0.5 TB).   
\end{itemize}

\begin{figure}

\begin{center}
\includegraphics[scale=0.35]{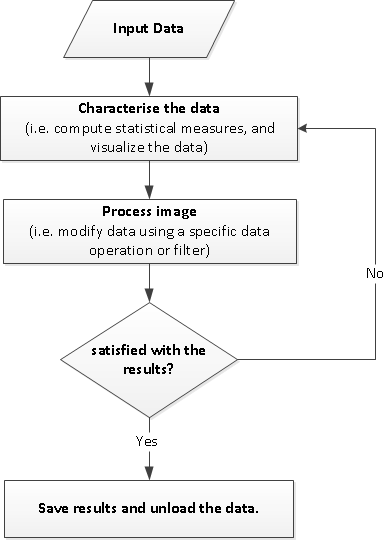}
\caption{Illustration for the new data analysis and processing model enabled by the presented framework.}
\label{fig:ProcessUpdate} 
\end{center}
\end{figure}

The presented framework suggests a new data analysis and processing model for the upcoming terascale data products. Data analysis, processing, and visualization work together to build computer-assisted iterative data analysis and processing pipeline. The steps of this processing model are described in Figure \ref{fig:ProcessUpdate}. The processing model starts by loading the data product and characterise it using methods such as computing statistical measures, and visualizing the data. Then, the data processing starts (e.g. modify data using a specific data operation or filter). The user can go through this interactively until a suitable outcome achieved.

The main objective of this data processing model is to reduce the I/O overhead by keeping the data in local memory and providing the user with an in-situ near realtime data analysis and visualization service. The current implemented data analysis and processing functions present a sample for the different features that can be added easily to the framework. The framework removes the burden of data management, processing elements synchronization, and results merging. The developer needs only to concentrate on ``how will I partition a task into smaller sub-tasks?'' and ``how should merging of these sub-tasks happen to ensure correctness?''.  

Due to restriction imposed by our test system, we were not able to use more than 96 GPUs within our benchmarks. With 96 GPUs, it is not possible to generate a balanced data partitioning [see \citet{hassan:2012a} for more technical details about the data partitioning mechanism]. That is why a special handling was added to the framework to support GPU numbers that are not a power of 2. This special handling allows us to load different data sizes (with total size less than the overall GPU memory) with no restrictions on the number of GPUs/nodes available but at the same time does not guarantee a balanced load between different GPUs. Given the reduction of the hardware cost and the relatively minor effect on the final user experience, sacrificing the balance between different GPUs is a reasonable and enabling compromise. 

Through our benchmarking on gSTAR we have shown that the framework can handle datasets over 0.5 terabyte with a frame rate better than seven frames per second. In general, the frame rate scales very well with the number of processing nodes, which means that the more processing units available, the better frame rate we can get until we reach the level where the communication overhead dominates the rendering time. This is shown more explicitly with the HIPASS dataset in Figure \ref{fig:HIPASSRenderingTime}. 

The performance difference between the generic transfer function implementation, and the maximum intensity projection is almost negligible in most cases. We still think the client-server processing mode \citep{hassan:2012a} is a better option to implement maximum intensity projection volume rendering, and will enable a better frame rate. 

The performance of the quantitative tools follows the same scalability pattern as for visualization tasks, which enables us to perform the data analysis processes discussed within this work within second rather than minutes. The framework can  perform tasks that require looping through $>$ 145 billion data points, such as calculating data mean and standard deviation, within less than two seconds (see Table \ref{tab:AnalysisSummary}). Even for sophisticated operation such as calculating the data median, the framework is capable of achieving that within 45 seconds for the 48X HIPASS dataset. As it is shown for datasets like the HIPASS dataset, this performance can be enhanced by increasing the number of computational units involved: from 13 seconds with eight GPUs down to two seconds with 96 GPUs. 

Having the GPUs as the main processing component, in addition to the communication and synchronization mechanism provided through this framework, is a principal element in achieving such scaling performance. Consequently, GPU properties including: the number of processing cores, the total amount of local memory, the communication bandwidth between the main system memory and the GPU local memory significantly affect  the system scalability and performance. 

The NVIDIA Fermi C2050 GPU model used to benchmark our framework has a theoretical processing power of 1.03 TFlops\footnote{\url{http://www.nvidia.com/docs/IO/43395/NV_DS_Tesla_C2050_C2070_jul10_lores.pdf}}, which is at least 5 times better than the latest Intel Sandy Bridge processor (up to 200 GFlops\footnote{\url{http://research.colfaxinternational.com/file.axd?file=2012\%2F4\%2FColfax\_FLOPS.pdf}} multi-threaded performance). If we assume perfect parallelism for a CPU similar solution, our framework will need around 480 Intel Sandy Bridge units, equivalent to 240 processing nodes, to achieve the same processing results. Such large increase ($\sim$ 2.5 times the number of nodes) in the number of processing nodes, given that the computational power is constant, will significantly increase the communication overhead and reduce the system performance. 

It is clear from Figure \ref{fig:HIPASSSampleOutput} that a regular screen resolution may not be the best output mechanism for tera-scale data cubes. A high resolution output will be preferable to achieve a better user experience from the qualitative and quantitative perspective. The current GUI client is capable of supporting different output resolutions, limited by the amount of GPU memory available after loading the dataset.

\section{Conclusions and Future work}

We present a case study of how qualitative and quantitative 3D visualization techniques, combined with the usage of high-performance computing and computational accelerators as the main processing infrastructure, can help astronomers to face the upcoming data avalanche. The presented framework is one of the first to enable in-situ data analysis and visualization for larger-than-memory astronomical data in near realtime. With the aid of GPUs as the main processing element, the framework raw performance approaches one teravoxel per second with only 48 nodes and 96 GPUs. We think the presented framework can enhance the current astronomical data analysis model by enabling computer-assisted iterative data analysis on massive datasets.

Radio astronomy was our main emphasis through this work due to the amount of data expected within the near future from different Square Kilometre Array pathfinder projects (e.g. ASKAP and the MeerKAT array), and ultimately the SKA itself. To our knowledge, the shown framework is the only available tool ready to support visualizing and analysing the expected output from SKA pathfinder projects. Additionally, this framework can be used with other different astronomical data products (e.g. cosmological simulation datasets - see \citet{hassan:2011a} for sample output). 

The framework can  support volume rendering for large astronomical data cubes with maximum intensity projection and generic transfer function as a qualitative data visualization tool. The framework also supports other quantitative data visualization techniques, including calculating minimum/maximum, calculating mean and standard deviation, computing a histogram, extracting 3D spectrum, and computing data median. The framework as well introduces a new sigma-clipping transfer function, which enables controlling the transfer function opacity based on the local noise properties.

The framework performance demonstrate the ability to render over 0.5 TB at better than seven frames per second. Within our benchmarks, we investigated the effect of increasing the output resolution and changing the number of processing nodes on the final framework performance. The framework has been shown to be scalable enough to handle larger datasets, provided appropriate hardware infrastructure is available. Additionally, quantitative data analysis processes have been benchmarked. The framework can  calculate simple data properties like the mean and standard deviation within less than two seconds for the 0.5 TB data cubes with 96 GPUs. Other more complicated data analysis task such as computing data median can be achieved in less than a minute.

Possible extensions to the current implementation that we are considering, include: (1) enabling multiple users to interact with the same dataset simultaneously, which will enable collaborative data analysis and visualization; and (2) handling time-dependent data (e.g Large Synoptic Survey Telescope output cubes and radio transient data), where the single dataset is relatively small but the number of cubes is large.

\section*{Acknowledgments}

This work was performed on the gSTAR national facility at Swinburne University of Technology. gSTAR is funded by Swinburne and the Australian Government’s Education Investment Fund.
We thank Dr. Jarrod Hurley (Swinburne University) and Gin Tan (Swinburne ITS) for their support with our usage for Swinburne gSTAR GPU Cluster. We thank Dr. Russell Jurek for providing sample data cubes. 

We used in our prototype implementation the following libraries : CFITSIO (\url{heasarc.gsfc.nasa.gov/fitsio/}), NVIDIA CUDA Driver API (\url{www.nvidia.com/cuda}), Clipper- polygon clipping library (\url{www.angusj.com/delphi/clipper.php}), and Snappy - compression/decompression library (\url{code.google.com/p/snappy/}).

\bibliographystyle{mn2e} 
\bibliography{references}

\appendix
\section[]{Torben's Median Method}
\label{App:TorbenMethod}

Algorithm \ref{Alg:Median} shows a pseudocode to describe the sequential implementation of the Torben's median method based on the ANSI C implementation by N. Devillard.

\onecolumn
\begin{algorithm}                      % enter the algorithm environment
\caption{Torben's Median Method}          % give the algorithm a caption
\label{Alg:Median}   
\begin{algorithmic}
\State \textbf{Input:} $Min,Max,data\_length$
\Loop
\State $Guess =\frac{(Min+Max)}{2}$
\State $LessCount = 0$
\State $GreaterCount = 0$
\State $EqualCount = 0$
\State $Max\_Less\_Than\_Guess = Min$
\State $Min\_Greater\_Than\_Guess = Max$

\For {$i=0$ to $data\_length$}
		\If {$Data[i]<Guess$}		
			\State $LessCount++$
			\If {$Data[i]>Max\_Less\_Than\_Guess$} 
				\State $Max\_Less\_Than\_Guess = Data[i]$
			\EndIf		 
		\ElsIf {$Data[i]>Guess$} 		
		.	\State $GreaterCount++$
			\If {$Data[i]<Min\_Greater\_Than\_Guess$} 
				\State $Min\_Greater\_Than\_Guess = Data[i]$
			\EndIf
		\Else 
			\State $EqualCount++$
		\EndIf	
\EndFor
\If {$(LessCount \leq \frac{(data\_length+1)}{2}) \;  and \;  (GreaterCount \leq \frac{(data\_length+1)}{2})$} 
	\State	$break$
\ElsIf {$LessCount>GreaterCount$} 
	\State $Max = Max\_Less\_Than\_Guess$ 
\Else 
	\State $Min = Min\_Greater\_Than\_Guess$
\EndIf

\EndLoop

\If {$LessCount \geq \frac{(data\_length+1)}{2}$} 
	\State $Median= Max\_Less\_Than\_Guess$
\ElsIf {$LessCount+EqualCount \geq \frac{(data\_length+1)}{2}$} 
	\State 	$Median= Guess$
\Else 
	\State 	$Median=Min\_Greater\_Than\_Guess$
\EndIf

\end{algorithmic}
\end{algorithm}

\bsp

%\label{lastpage}

\end{document}